\documentclass[12pt]{iopart}
\usepackage[sort&compress, numbers]{natbib}
\usepackage{tabularx}
\usepackage[utf8]{inputenc}
\usepackage{subcaption}
\usepackage{comment}
\usepackage{bbold} 
\usepackage{hhline}
\usepackage{braket}
\usepackage{balance}
\usepackage{physics}
\usepackage{dcolumn}
\usepackage{bm}
\usepackage[unicode=true,bookmarks=true,bookmarksnumbered=false,bookmarksopen=false,breaklinks=false,pdfborder={0 0 1},backref=false,colorlinks=true,linkcolor=cyan,citecolor=cyan]{hyperref}
\usepackage{mathdots}
\usepackage[]{graphics,graphicx,epsfig}
\usepackage{mathtools}
\usepackage{csquotes}
\MakeOuterQuote{"}
\usepackage{epstopdf}

\def\bea{\begin{eqnarray}}
\def\eea{\end{eqnarray}}

\usepackage[x11names]{xcolor}

\DeclarePairedDelimiter{\floor}{\lfloor}{\rfloor}

\begin{document}

\title{Integer Programming Using A Single Atom}

\author{Kapil Goswami$^{1,*}$, Peter Schmelcher$^{1,2}$ and Rick Mukherjee$^{1,\dag}$} 

\address{%
 $^1$Zentrum f\"ur Optische Quantentechnologien, Universit\"at Hamburg, Luruper Chaussee 149, 22761 Hamburg, Germany
}%
\address{%
 $^2$The Hamburg Centre for Ultrafast Imaging, Universit\"at Hamburg, Luruper Chaussee 149, 22761 Hamburg, Germany
}%
\ead{\mailto{$^*$kgoswami@physnet.uni-hamburg.de} and \mailto{$^{\dag}$rick.mukherjee@physnet.uni-hamburg.de}}

%\vspace{10pt}
%\date{\today}
\begin{abstract}
Integer programming (IP), as the name suggests is an integer-variable-based approach commonly used to formulate real-world optimization problems with constraints. Currently, quantum algorithms reformulate the IP into an unconstrained form through the use of binary variables, which is an indirect and resource-consuming way of solving it.
We develop an algorithm that maps and solves an IP problem in its original form to any quantum system possessing a large number of accessible internal degrees of freedom that are controlled with sufficient accuracy.
This work leverages the principle of superposition to solve the optimization problem.
Using a single Rydberg atom as an example, we associate the integer values to electronic states belonging to different manifolds and implement a selective superposition of different states to solve the full IP problem.
The optimal solution is found within a few microseconds for prototypical IP problems with up to eight variables and four constraints. This also includes non-linear IP problems, which are usually harder to solve with classical algorithms when compared to their linear counterparts.
Our algorithm for solving IP is benchmarked by a well-known classical algorithm (branch and bound) in terms of the number of steps needed for convergence to the solution. This approach carries the potential to improve the solutions obtained for larger-size problems using hybrid quantum-classical algorithms.
\end{abstract}
%
% Uncomment for keywords
\vspace{2pc}
\noindent{\it Keywords}: real-world optimization, integer programming, single atom, superposition principle, quantum system, optimal control 
%
% Uncomment for Submitted to journal title message
%\submitto{\JPA}
%
% Uncomment if a separate title page is required
%\maketitle
% 
% For two-column output uncomment the next line and choose [10pt] rather than [12pt] in the \documentclass declaration
%\ioptwocol
%
\maketitle
\section{Introduction}
\label{introduction}
\begin{figure}[t]
\centering
\includegraphics[width = 0.6\textwidth,trim={20cm 9cm 18cm 9cm},clip]{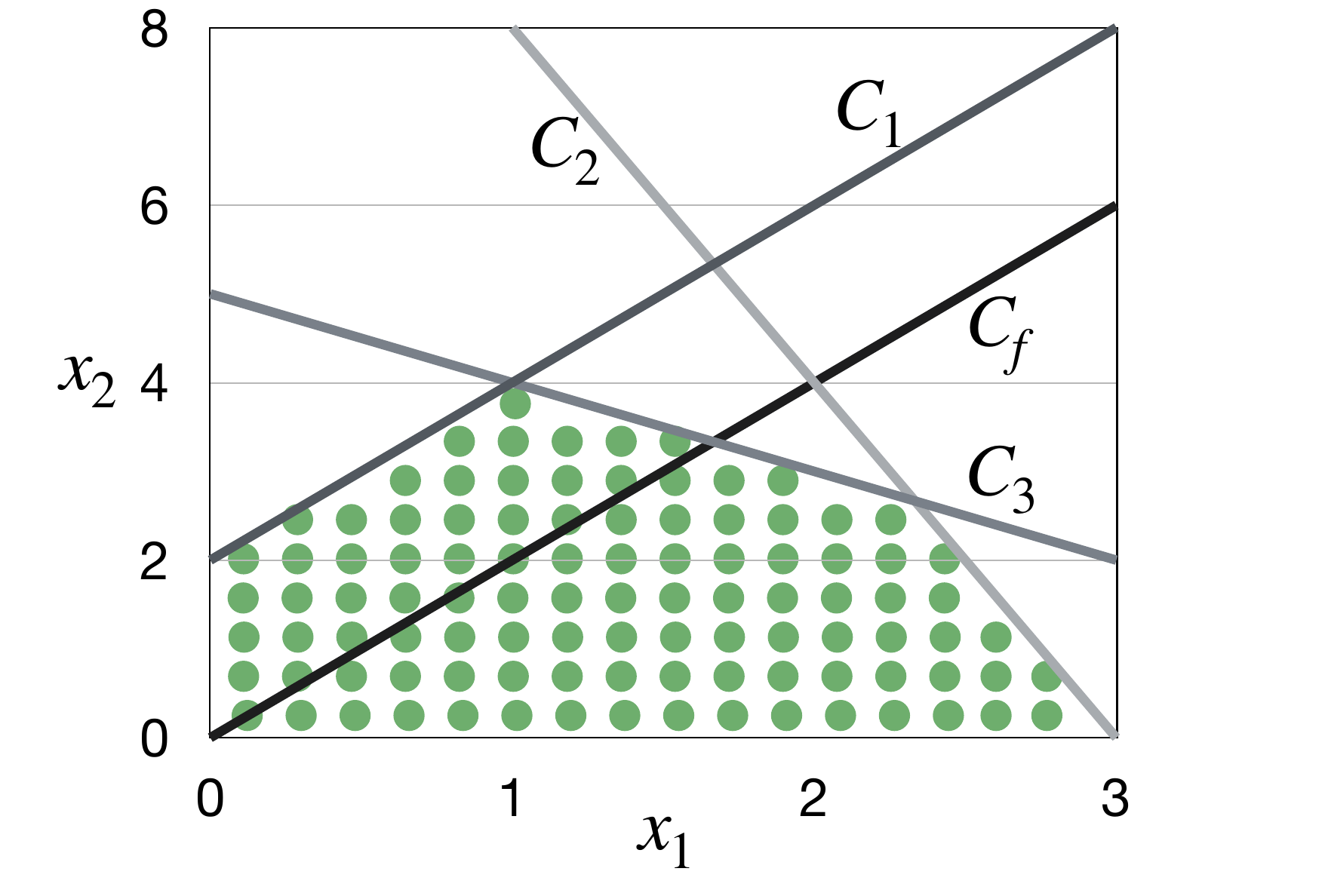}
\caption{The schematic plot represents a typical integer programming problem consisting of two variables $x_1$ and $x_2$ with three constraints ($C_1$, $C_2$, and $C_3$). The cost function $C_f$ is maximized to solve the problem while satisfying the constraints. The dotted region depicts the allowed solution space (feasible region)}.
\label{ILP}
\end{figure}

Quantum algorithms have witnessed significant milestones in the past, for instance, Shor's algorithm for factoring large numbers, Grover's algorithm for searching an unsorted database, and the Deustch-Jozsa algorithm for finding whether a function is a constant or balanced \cite{montanaro2016quantum}.   
While these algorithms show theoretical quantum advantages, their realization faces practical challenges. This is due to the requirement of a large number of noiseless qubits that lie outside the capabilities of current quantum devices \cite{bharti2022noisy}. 
Quantum algorithms that solve optimization problems \textit{efficiently} can have an immediate impact on industry-related applications and offer a practical advantage. 
Many of the real-world optimization problems contain discrete variables such as different cities in the traveling salesman problem \cite{matai2010traveling}, zones in energy market operation \cite{alizadeh2011reliability} and, number of days in production planning \cite{pochet2006production}. These problems are widely formulated in terms of integer programming (IP) \cite{wolsey2020integer} which is a variable assignment problem in which the cost function is optimized under given constraints and each variable can take integer values as shown in Figure~\ref{ILP}. Generalizations of these problems are known as mixed-integer programming (MIP) where variables are both integers as well as of continuous type. Solving IP as well as MIP problems lie in the computational complexity class of NP-hard \cite{papadimitriou1982complexity,papadimitriou1988optimization, kannan1978computational} and solving them is an ongoing research area \cite{wolsey2020integer,wolsey2007mixed,bliek1u2014solving}. The complexity arises from the combinatorial aspect of the problem due to integer variables. If all the variables in MIP are continuous, then the problem can be solved in polynomial time using a classical computer \cite{schrijver1998theory}.
The integer variables serve as a bottleneck for tackling optimization problems classically which makes IP an ideal candidate for benchmarking quantum algorithms to establish an advantage over classical algorithms.

Developing an algorithm that can be implemented directly on a quantum device to solve IP remains an open challenge.
The few examples of quantum algorithms in the literature that solve IP problems on a quantum system use an indirect mapping of integer to binary variables, and converting the problem to an unconstrained form \cite{chang2020hybrid,okada2019efficient,svensson2023hybrid,bernal2020quantum,ajagekar2022hybrid,bernal2020integer,khosravi2023mixed}. 
The choice of binary variables stems from the fact that they mimic interacting spins that are easily implementable on current quantum simulators and gate-based quantum computers \cite{saffman2010quantum,shi2022quantum,bruzewicz2019trapped,goswami2023solving,henriet2020quantum}.
Since the variables associated with most real-world problems naturally map to integers instead of binary variables, the encoding of the IP problem becomes resource-consuming \cite{bernal2020quantum}. 
A quantum algorithm in principle is a set of instructions that run on a quantum system where it leverages concepts like the superposition principle and/or entanglement to achieve a quantum advantage. 
It is not clear whether entanglement is a necessary pre-requisite for speed-up over classical algorithms \cite{biham2004quantum,kenigsberg2006quantum,lloyd1999quantum,lanyon2008experimental}. In certain cases, it has been argued that the principle of superposition is sufficient to get a polynomial speed-up \cite{kenigsberg2006quantum,lloyd1999quantum}.
In this work, we introduce a direct method to encode the integer variables and solve any IP problem by exploiting the superposition of multiple accessible degrees of freedom of a quantum system.

\begin{figure*}[t]
\centering
\includegraphics[width = 1\textwidth,trim={0cm 7cm 15cm 4.5cm},clip]{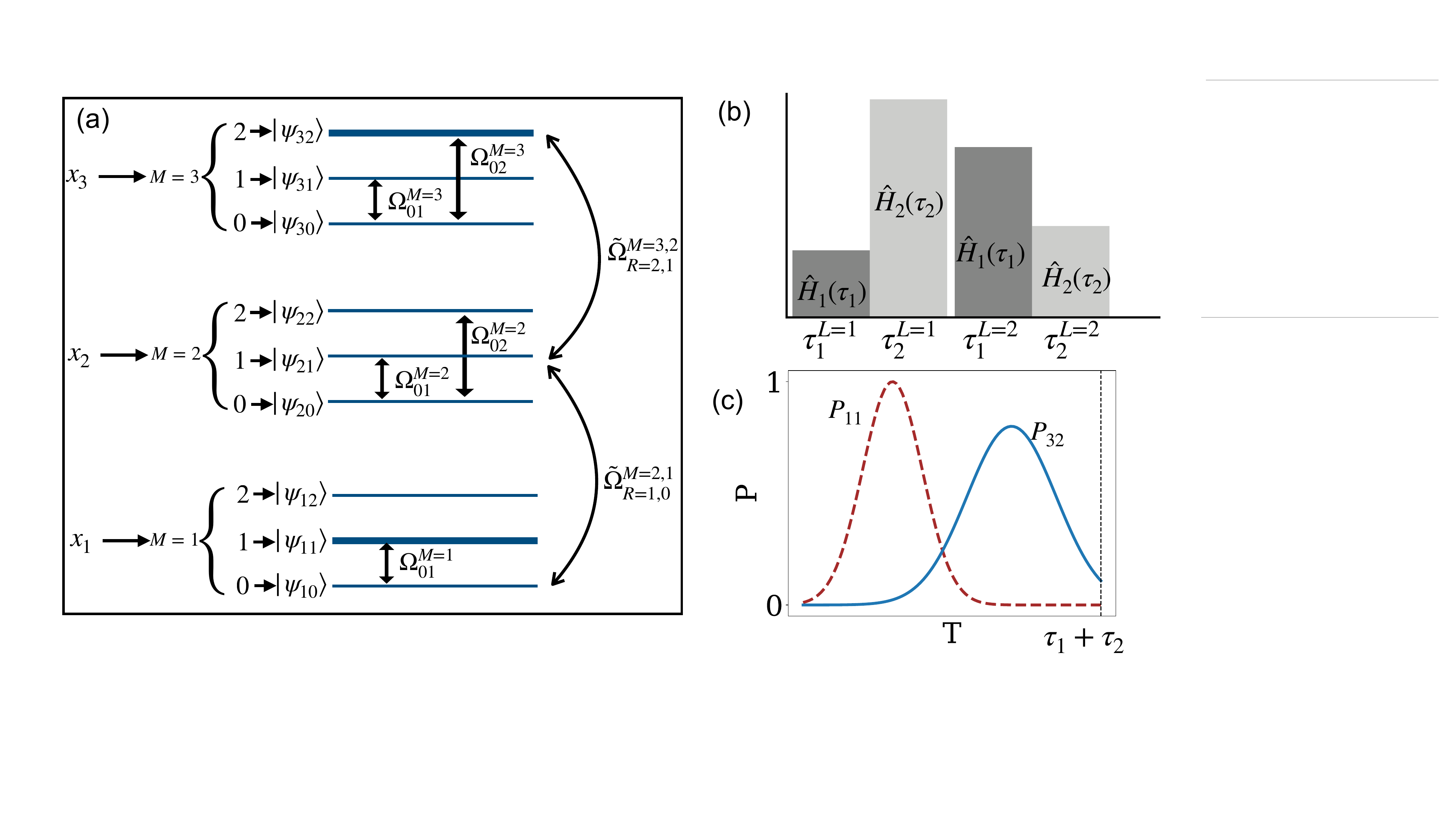}
\caption{The Figure summarizes the algorithm for solving IP problems using a multi-level system. (a) As explained in the text, the three manifolds correspond to three individual integer variables. The external couplings $\tilde{\Omega}$ are between the manifolds $M=1,2,3$ while the internal couplings $\Omega$ are between the states ($\ket{\psi_{i0}},\ket{\psi_{i1}},\ket{\psi_{i2}}$) residing in a particular manifold $M=i$. The value that each variable can take is associated with the population of the dominant internal level belonging to a manifold. (b) Schematically shows the coupling Hamiltonians $\hat{H_1}(\tau_1)$ and $\hat{H_2}(\tau_2)$ used to evolve an initial state, used for describing constraints as shown in Equations~(\ref{H1eq}-\ref{H2eq}). The height of the shaded rectangle indicates the strength of the coupling while the width is the evolution time, both of which are varied in the optimization procedure. During the optimization, the coupling Hamiltonians are applied repeatedly in layers indicated by $L$. (c) Shows the population transfer from one state to another during the first layer of the protocol in time $\tau_1+\tau_2$. $P_{11}$ and $P_{32}$ correspond to the population of the states $\psi_{11}$ and $\psi_{32}$ (shown as states with thick blue lines in panel a) respectively, assigning the values to the variables as $x_1=1$ and $x_3=2$.} 
\label{Setup}
\end{figure*}

Examples of quantum systems with a large number of controllable degrees of freedom include hyperfine/electronic states in atoms \cite{kanungo2022realizing,gadway2015atom} or non-trivial superposition of them like Rydberg-dressed states \cite{macri2014rydberg,mukherjee2019charge}, frequency modes in photonic system \cite{ozawa2019topological,yuan2018synthetic}, rotational states in ultracold polar molecules \cite{shaffer2018ultracold,sawant2020ultracold,gadway2016strongly} and the simulation of \textit{synthetic} dimensions \cite{boada2012quantum,martin2017topological,sundar2018synthetic}. Any of these systems mentioned above can be used to implement our scheme for encoding and solving IP problems, however, in this work, the focus is on a single atom with multiple manifolds of states. 
Each manifold corresponds to a variable in the problem while the levels in each manifold are assigned different integer values the variable can take. 
The manifolds and the states are coupled through lasers to implement certain aspects of the constraints as shown in Figure~\ref{Setup}(a). These couplings are optimized in time and selectively transfer the population of the states to find the integer variables that are the solution to the IP problem, refer to Figure~\ref{Setup}(b),(c). 
This scheme applies to a wide range of IP problems with different levels of complexity, generally characterized by the number of non-linear terms, types of constraints, and size of the problem.

This manuscript is structured as follows, the Theory Section \ref{theory} provides a brief description of the mathematical problem of IP and the steps for implementing an algorithm to solve it. 
The algorithm is detailed using a Rydberg atom as an exemplary case.  This is followed by a discussion on the complexity of IP problems which determines the choice of the problems we solve.
The Results Section \ref{results} highlights the key feature of our mapping for certain chosen IP problems and a benchmarking result with a classical algorithm. Section \ref{Conclusion} contains our conclusions and outlook.

\section{Theory and algorithmic implementation}
\label{theory}
\begin{comment}
We introduce the integer programming (IP) problem in subsection \ref{IP}. 
A mapping of a general IP problem to the energy levels of an atom and the steps of the algorithm is provided in subsection \ref{QA}.
The algorithm in subsection \ref{Rydberg} is presented using the specific physical settings of a Rydberg atom.
The complexity of the different types of IP problems is discussed by identifying relevant benchmarking metrics in subsection \ref{Complexity}.
\end{comment}
\subsection{Integer programming}
\label{IP}

An integer programming (IP) problem is an optimization problem, such that the decision variables are constrained to take integer values \cite{wolsey2020integer}. A typical IP problem is depicted in Figure~\ref{ILP}, the problem has two variables $x_1,x_2$ and three constraints $C_1,C_2,C_3$. The highlighted area is called a \textit{feasible region} as all the constraints are satisfied, and the cost function $C_f$ is optimized to find the extremal in the feasible region. 
Consider $m$ optimizing variables $\mathbf{x}:= (x_1, ..., x_m)$ with $m'$ terms in the cost function with coefficients $\mathbf{c}:= (c_1,...,c_m')$, $A$ be a $n \times m$ matrix describing $n$ constraints, and the condition on the constraints given by $\mathbf{b}:= (b_1,...,b_n)$.
The integer programming problem to maximize the cost function $C_f(\mathbf{x})$ can be represented as
    \begin{equation}
        C_f(\mathbf{x}) = c_1 x_{p=1}\cdots x_{p=m} + \cdots + c_{m'} x_{p'=1} \cdots x_{p'=m},
    \end{equation}
where $p, p' \in \mathbb{Z}^+$ and $x_i \in \mathbb{Z}$, under the constraints given as
\begin{align}
    \begin{aligned}
    C_1: a_{11}x_{q=1}\cdots x_{q=m} +\cdots  + a_{1m}x_{q'=1}\cdots x_{q'=m} \leq b_1 \\
             \vdots & \\
        C_n:a_{nm}x_{r=1}\cdots x_{r=m} + \cdots  + a_{nm}x_{r'=1}\cdots x_{r'=m} \leq b_n
    \end{aligned}
\end{align}

where $q,q',r,r' \in \mathbb{Z}^+$. If the objective function and the constraints in the problem are linear, then the problem is termed integer linear programming (ILP) otherwise it falls into the non-linear integer programming (NLIP) category. Specifically, $C_f,C_1,...,C_n$ all have to be linear for the problem to be called ILP. 
Without loss of generality, constraints of the form $C_1,...,C_n$ cover all the cases of an IP problem.

Multiple classical solvers are available to solve IP such as GUROBI, IBM CPLEX, SCIP, Xpress, GLPK, MOSEK, BARON \cite{meindl2012analysis,andersen2000mosek,kilincc2018exploiting}. The underlying algorithms for many of the classical solvers can be broadly categorized as Branch and Bound (B\&B)-based or simplex-based. Branch and Bound (B\&B) \cite{beale1965mixed} involves breaking down the problem into smaller subproblems, solving them individually, and using bounds to reduce the search space efficiently. Branch and Cut is an extension of the B\&B method that incorporates cutting planes to tighten the bounds on the solution space \cite{sen2006decomposition}. B\&B-based solvers run into issues if the relative continuous relaxation gap increases for a problem, requiring a large number of iterations to converge \cite{kronqvist2019review}. In comparison, simplex-based solvers are mostly used to solve linear problems. Any non-linear problem is first converted into a linear one for the simplex algorithm application, and the whole process becomes computationally expensive. Both, B\&B-based and simplex-based solvers have advantages and shortcomings based on the type of IP problem that is being solved. Currently, there are no quantum schemes that leverage the inherent structure of any quantum system to \textit{efficiently} encode and solve IP problems, except for translating the problem to quadratic unconstrained binary optimization (QUBO) \cite{glover2019quantum} form.

\subsection{An algorithm for solving integer programming}
\label{QA}

To demonstrate the fundamental principles of our approach, a sample problem with two constraints $C_1,C_2$, cost function $C_f(\mathbf{x})$ to be maximized, and three variables $x_{i=1,2,3} \in \{0,1,2\}$ is considered, 

\begin{equation}
\begin{split}
C_f(\mathbf{x}) =3 x_1 + 2x_2 +  x_3 \\
C_1: 2x_1+x_2 \leq 3 \\
C_2:x_2+x_3 \leq 2.
\end{split}
\label{ExP}
\end{equation}

The proposed quantum algorithm consists of four steps, (1) Mapping of variables onto internal states of a quantum system (2) Mapping of the constraints of the problem through couplings of the levels (3) Optimization of the objective function for solving the problem (4) Measurement of the state population. This will be followed up using the Rydberg atom as an example. \\

\noindent \textbf{Step 1: Mapping of variables onto internal states} -
The variables in the IP problem $x_{i=1,2,3}$ are one-to-one mapped to different energy manifolds $M=1,2,3$ consisting of multiple states as shown in Figure~\ref{Setup}(a). The variable $x_i \in \{0,1,2\}$ corresponds to each manifold $M=i$ consisting of three states $\ket{\psi_{i0}},\ket{\psi_{i1}}$ and $\ket{\psi_{i2}}$ with values assigned as $ 0, 1$ and $2$ respectively. 
Depending on the problem, the variables can take any integer values which then determines the number of sub-levels in a manifold. Naturally, this is restricted by the allowed number of physically accessible internal states in the quantum system.
The population of the states within each manifold is measured to decode the value of the corresponding variables. At the end of the protocol, the state with the highest population in the manifold determines the integer value of that particular variable. 
As an example, $x_1$ represented by the first manifold $M=1$ in Figure~\ref{Setup}(a), takes the value $1$ if the $\ket{\psi_{11}}$ is the highest populated level in manifold $M=1$.   
Similarly, in Figure~\ref{Setup}(c) population of the states $\ket{\psi_{11}},\ket{\psi_{32}}$ is measured to be dominant, corresponding to the value of variables $x_1=1, x_3=2$ respectively.\\

\noindent \textbf{Step 2: Mapping of the constraints} - To implement the constraint $C_i$ physically, the key idea is to couple the states of a manifold (by internal couplings) corresponding to the allowed values of that particular variable and externally couple the states in different manifolds (termed as external couplings) to establish the relationship between the variables for a given constraint. 
Specifically, the internal couplings $\Omega_{ij}^{M}$ describe the coupling of the internal states of $M$ that are used within a manifold to restrict populating specific states that are excluded by the problem constraint. 
The external couplings $\tilde{\Omega}_{M, i}^{R,j}$ on the other hand, are between the states of different manifolds $M, R$ to implement the relationship between the variables. By tuning $\tilde{\Omega}_{M, i}^{R,j}$, the population of the levels in different manifolds redistributes and in turn couples the corresponding variables.
Both the couplings $\Omega_{ij}^{M}(t)$, $\tilde{\Omega}_{M, i}^{R,j}(t)$ are time-dependent to facilitate a dynamic population transfer between the levels and provide controls that can be optimized to reach a configuration corresponding to the inequality being satisfied. 
The corresponding time-dependent Hamiltonian is given by, 

\begin{equation}
\begin{split}
\hat{H}(t) = \sum_{M,k,R,l} (\tilde{\Omega}^{M,l}_{R,k}(t) \ket{\psi_{Rk}}\bra{\psi_{Ml}} + h.c.) 
+
\sum_{M,i,j} (\Omega^{M}_{ij}(t) \ket{\psi_{Mi}}\bra{\psi_{Mj}} + h.c.),
\end{split}
\label{Heq}
\end{equation} 

where $\ket{\psi_{Mk}}$ is the state $k$ in manifold $M$. 
The coupling of state $\ket{\psi_{Rk}}$ belonging to a manifold $R$ to the state $\ket{\psi_{Ml}}$ in the other manifold $M$ is described by the first term in $\hat{H}(t)$. In the second term, the constraints for individual variables are given by the coupling of the states $\ket{\psi_{Mi}}$ to $\ket{\psi_{Mj}}$ within the same manifold $M$. 
For a given problem, the number of required internal and external couplings are predefined and hence many of the terms in $\hat{H}(t)$ will vanish.
The steps outlined for encoding the problem apply to any IP, both for linear and non-linear cases.
In general, multiple inequalities need to be satisfied, which requires a coupling Hamiltonian $\hat{H_{i}}(t)$ for each constraint $C_i$. However, the inequalities must be fulfilled simultaneously to solve the problem.

For example, let's consider an example problem given by Equation~(\ref{ExP}) with three variables and two constraints $C_1,C_2$. 
Any implementation of the constraints on the quantum system is divided into three parts, restricting the values of the individual variables, constructing the relationship between the variables, and 
satisfying the inequalities. 
In the constraint $C_1$, $x_{1} = 2$ is not allowed which translates to only coupling $\ket{\psi_{10}}$ and $\ket{\psi_{11}}$ states in the manifold $M = 1$ which avoids populating the state $\ket{\psi_{12}}$. 
The coupling term $\Omega_{01}^{M=1}$ as shown in Figure~\ref{Setup}(a) is used for associating selective states. As for variable $x_2$, both $\Omega_{01}^{M=2}$ and $\Omega_{02}^{M=2}$ are non-zero as it can take any value in $\{0,1,2\}$.
The variables $x_1$ and $x_2$ in $C_1$ are linked by the external coupling $\tilde{\Omega}_{M=2, i}^{R=1,j}$ which connects the manifolds $M=2$ and $R=1$. 
Similarly for another constraint $C_2:x_2 + x_3 \leq 2$, $\tilde{\Omega}_{M=2, i}^{R=3,j}$ will couple $M=2$ and $R=3$ for the mapping. 
The allowed couplings will depend on the states chosen in each manifold as shown schematically in Figure~\ref{Setup}(a).
The individual Hamiltonians derived from Equation~(\ref{Heq}) for the example problem (constraints $C_1$ and $C_2$) are given as,

\begin{equation}
\begin{split}
\hat{H_1}(t) = \tilde{\Omega}^{M=1,0}_{R=2,1}(t) \ket{\psi_{10}}\bra{\psi_{21}}
+ \Omega^{M=1}_{01}(t) \ket{\psi_{10}}\bra{\psi_{11}}  +  \Omega^{M=2}_{01}(t) \ket{\psi_{20}}\bra{\psi_{21}} \\+ \Omega^{M=2}_{02}(t) \ket{\psi_{20}}\bra{\psi_{22}} + h.c.,
\end{split}
\label{H1eq}
\end{equation} 

\begin{equation}
\begin{split}
\hat{H_2}(t) = \tilde{\Omega}^{M=2,1}_{R=3,2}(t) \ket{\psi_{21}}\bra{\psi_{32}} +  \Omega^{M=2}_{01}(t) \ket{\psi_{20}}\bra{\psi_{21}}  + \Omega^{M=2}_{02}(t) \ket{\psi_{20}}\bra{\psi_{22}} \\
+  \Omega^{M=3}_{01}(t) \ket{\psi_{30}}\bra{\psi_{31}}  
+ \Omega^{M=3}_{02}(t) \ket{\psi_{30}}\bra{\psi_{32}} + h.c..
\end{split}
\label{H2eq}
\end{equation} 

An initial state $\ket{\psi_{initial}}$ is chosen with a population of the state $\ket{\psi_{10}}$ be $1$. Let $p_{ij}$ describes the population of the state $\ket{\psi_{ij}}$ and the initial population  distribution is given as $p_{10}=1, p_{11}=0, ..., p_{32}=0$. Then we apply $\hat{H_{1}}(t)$ for time $\tau_1$ and $\hat{H_{2}}(t)$ for time $\tau_2$ to evolve the initial state as $\hat{H_{2}}(\tau_2) \hat{H_{1}}(\tau_1) \ket{\psi_{initial}}$, which describes a layer $L$, shown in Figure~\ref{Setup}(b). 
The layering helps in population mixing and multiple such layers (with the same Hamiltonians) are constructed with the number of layers $L$ being predefined for each problem. The choice of the coupling states, time $\tau$, and the strength of the couplings are to be optimized to fulfill all the constraints simultaneously while finding the maximal cost function, which is described in the next step. \\

\noindent \textbf{Step 3: Optimization of the objective function} -  
Typically, the quantum optimal control requires the initialization of the iteratively varied parameters to find the extrema of an objective function. In our case, the coupling strengths $\Omega(t),\tilde{\Omega}(t)$ and the time $\tau$ corresponding to each Hamiltonian (like in Equations~(\ref{H1eq}-\ref{H2eq}) and Figure~\ref{Setup}(b)) are initially chosen to an arbitrary value. Through an iterative process, optimal values of these parameters are found using an optimizer, and the results for these optimal parameters are shown in Section \ref{results}, where different prototypical problems are solved. The sum of all the values for $\tau$ corresponds to the total time $T$ of the optimal protocol, where $T$ is not fixed but is upper-bounded by the lifetime of the quantum system. 
The variation in the couplings with time~$\tau$ results in the selective population transfer of the states.
The integer variables are decoded from the population of the manifolds, and then the cost function that needs to be optimized is calculated. We can determine the time intervals for which all the inequalities are satisfied and form a set $S_{\tau}$ consisting of those time points, for example, the shaded region in Figure~\ref{QILP1}(a). The goal is to minimize the objective function $O$, which is calculated from the variables decoded from the population. 
The objective function contains two parts, satisfying all the constraints simultaneously and maximizing the cost function $C_f$ of the problem \cite{kuros2021controlled,krauss2023optimizing}. 
The multi-objective function is defined as 
    \begin{equation}
    O|_{min} = (1- \frac{N_{\tau}}{N_T})+ (1-\frac{\sum_{t \in S_{\tau}}C_f (t)}{\sum_{t=0}^{t=T} C_f (t)}),
    \label{OF}
    \end{equation}
where $T$ is the total time of the protocol, $N_T$ is the total number of discretized time points of $T$, $S_{\tau}$ is a set containing time points where all the inequalities/constraints are satisfied, $N_{\tau}$ is the cardinality of the set $S_{\tau}$, and $C_f(t)$ is the cost function that needs to be maximized in the integer programming problem. 
The first term of the objective function maximizes $N_{\tau}$ i.e. the total time for which all the constraints are simultaneously satisfied, ensuring that the system remains in the feasible region of the IP problem for a longer duration.  
The second term maximizes the sum of the cost function at all times in the feasible region ($t\in S_{\tau}$). This increases the probability of finding the optimal solution during the measurement in an experiment and contributes to the robustness of the protocol.  
It is essential to normalize each term in $O$ to prevent bias toward one of the objectives which can result in skewed-inaccurate results.
The multi-objective function is then minimized using a combination of a gradient (BFGS) \cite{broyden1970convergence,fletcher1970new,goldfarb1970family,shanno1970conditioning} and non-gradient (Nelder-mead) \cite{nelder1965simplex} based optimal control methods to find the optimal values of the couplings. There exist other quantum optimal control methods such as Bayesian optimization that can also be used \cite{mukherjee2020bayesian,mukherjee2020preparation}. \\

\noindent \textbf{Step 4: Measurement of the state population} -  
During the optimization of couplings, the protocol needs to provide time intervals where all constraints are simultaneously satisfied while maximizing the cost function. Such optimization requires us to arbitrarily choose the time intervals for regular measurement, which is set at every microsecond. The population of the levels is measured during these intervals to decode the values of the variables that yield the optimal solution.
If the solutions are obtained within $30\mu s$, this necessitates performing 30 population measurements, each potentially requiring multiple repetitions depending on the experimental setup.

\subsection{Physical implementation in a Rydberg atom}
\label{Rydberg}
Selectively populating and creating a superposition of states of a multi-level system allows us to implement IP problems under the above framework. In principle, any quantum system can be used for implementing our algorithm. However, inspired by recent experiments such as \cite{kanungo2022realizing}, here we provide the steps of our protocol for a single Rydberg atom.

A Rydberg atom with the principal quantum numbers $n = 56,...,62$, and angular momentum $l=0,1$ is considered for the simulations \cite{kanungo2022realizing}. Initial state preparation involves a population transfer from the atom's ground state to one of the Rydberg states participating in the multi-level transitions. For the example problem given by Equation~(\ref{ExP}), each manifold consists of 3 states corresponding to the constraint on each variable $x_i~\in~\{0,1,2\}$. To allow maximum flexibility in terms of transitions, the Rydberg states in one of the manifolds are chosen to be $\ket{nS}, \ket{nP}$, and $\ket{(n+1)P}$. The next manifold representing another variable, contains the states $\ket{n'S}$, $\ket{n'P}$ and $\ket{(n'+1)P}$, such that $n \neq n'$ and $n \neq n'+1$. The following assignment for the Rydberg states can be used for applying the algorithm to the example problem.
\begin{equation}
\begin{split}
    \ket{\psi_{10}} \xrightarrow{} \ket{nS}, \ket{\psi_{11}} \xrightarrow{} \ket{nP}, \ket{\psi_{12}} \xrightarrow{} \ket{(n+1)P} \\
    \ket{\psi_{20}} \xrightarrow{} \ket{n'S}, \ket{\psi_{21}} \xrightarrow{} \ket{n'P}, \ket{\psi_{22}} \xrightarrow{} \ket{(n'+1)P} \\
    \ket{\psi_{30}} \xrightarrow{} \ket{n''S}, \ket{\psi_{31}} \xrightarrow{} \ket{n''P}, \ket{\psi_{32}} \xrightarrow{} \ket{(n''+1)P} \\ 
\end{split}
\end{equation}

The Hamiltonian for the coupling of the states in a single Rydberg atom with detuning $\Delta = 0$ is given by Equation~(\ref{Heq}), thereby neglecting the counter-rotating terms in the couplings, transforming into a rotating frame and absorbing the constants in the couplings. The states are coupled through microwave lasers with Rabi frequencies $\Omega$ and $\tilde{\Omega}$. Selectively populating Rydberg atoms is analogous to having a multilevel extension of the electromagnetically induced transparency process \cite{mohapatra2007coherent}. Other ways to control the state populations can involve the use of stimulated Raman adiabatic passage (STIRAP) \cite{cubel2005coherent}. By employing optimal quantum control methods, similar to our algorithm, other approaches can be used to efficiently populate the states \cite{kanungo2022realizing}. 

After populating the states for solving a given problem, the task is to make repeated measurements to efficiently decode the values of the variables. The measurement of the populations can either be done by selective field ionization ($\mu s$ time scales) \cite{gallagher1977field} or time-resolved measurements ($ns- \mu s$ time scales) \cite{cetina2016ultrafast}. Particularly, for measuring the population of the states in the Rydberg atoms, selective field ionization can be performed with tens of $V/cm$ electric field strength and within a time scale of a few $\mu s$ \cite{kanungo2022realizing}. The time it takes for the measurement can be brought down by increasing the electric field strength or by choosing a different principal quantum number $n$ for the Rydberg states. The selective field ionization is a destructive process of imaging. Thus repeated experiments are needed to measure the state population under the assumption that the optimal sequence of pulses
produces the same desired population in an experiment where the noise is under check as was the case in \cite{kanungo2022realizing}. The typical lifetime for a Rydberg state is much larger than the time required to reach the solution ($\sim 30 \mu s$) using our algorithm. For comparison, the lifetime for a Cesium atom in $\ket{65D_{5/2}}$ state is $\sim 94 \mu s$ \cite{song2022lifetime} and this lifetime also increases with the principle quantum number $n$. The results shown in Section \ref{results} are well within the reach of the current experimental capabilities.

\subsection{Complexity of the problem}
\label{Complexity}
We now discuss the computational complexity and the difficulty of solving different instances of the problem.
Linear programming (LP) problems (without integer constraints) transition from the P-complexity class to mixed-integer programming (MIP) problems (with integer constraints) that become NP-hard due to the added complexity of exploring discrete solution spaces.
The NP-hardness of the problems can be classified in a hierarchy as BILP $\subset$ ILP $\subset$ MILP $\subset$ MNLIP, where B is binary, L stands for linear and NL stands for non-linear. 
Integer constraints always make the problem non-convex \cite{wolsey1999integer}, which results in standard convex optimization techniques that work efficiently for continuous problems no longer being applicable. However, in this work, we refer to the convexity (or non-convexity) of any problem as a geometric property of the \textit{feasible region}, determined by converting the discrete problem to a continuous one. The constraints describe the boundaries of the \textit{feasible region} and if the region enclosed has a convex boundary then the problem is called convex. The objective function is then maximized or minimized in that region. If the constraints do not enclose any overlapping region then the problem becomes infeasible as all the constraints cannot be simultaneously satisfied. However, the concavity in the \textit{feasible region} can encompass multiple local optima, saddle points, or discontinuities, making it harder for the standard solvers to navigate the optimization landscape. 
There is a special class of IP problems with a totally unimodular constraint matrix that does not fall into the category of NP-hard \cite{schrijver1998theory}. This shows that the complexity of the problems can also vary from one instance to another, one of the trivial metrics is the size of the NP-hard problem. In this Section, three parameters \cite{kronqvist2019review} are discussed that can be used to encapsulate the difference in the difficulty of solving a general IP problem. 
\begin{figure*}[t]
\centering
\includegraphics[width = 1\textwidth,trim={0cm 0cm 0cm 0cm},clip]{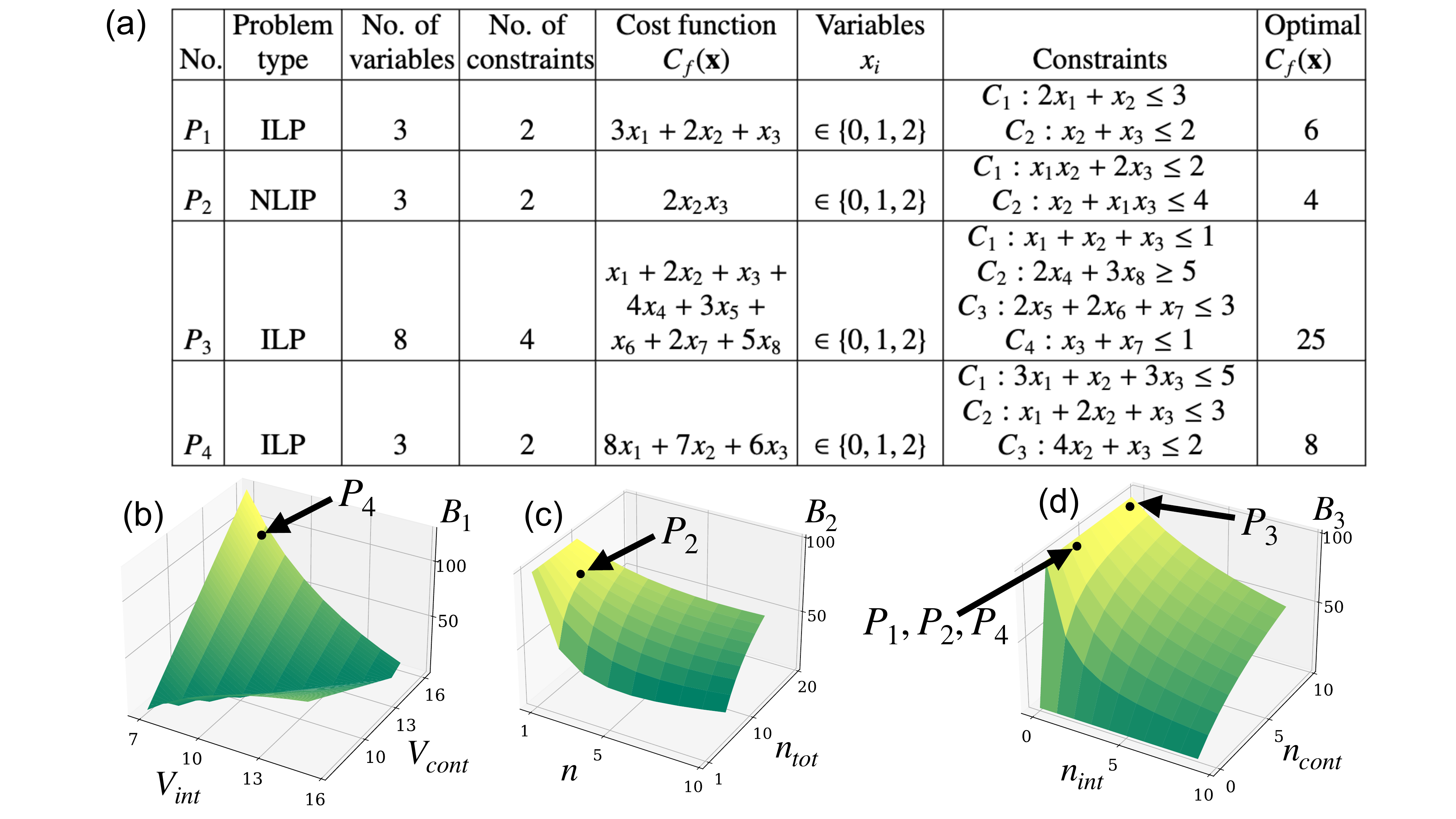}
\caption{The table in (a) shows the chosen sample integer programming problems that are encoded and solved using multi-levels of a single Rydberg atom. The problems with varied complexity (as explained in Sub-section \ref{Complexity}) depend on the problem type and the number of variables. The corresponding benchmark metrics \cite{kronqvist2019review} are shown in (b-d) panels, where the black dot indicates the complexity of the problems $P_1-P_4$ as defined in panel (a).}
\label{Tab}
\end{figure*}

The relative continuous relaxation gap is one of the benchmark metrics that captures the difference between the optimal solution of the IP problem and the relaxed IP problem, defined as
\begin{equation}
B_1 = \frac{|V_{int}-V_{cont}|}{max\{|V_{int}|,0.001\}} \cross 100\%,
\label{B1}
\end{equation}
where $V_{int}$ is the optimal value of the integer problem and $V_{cont}$ is the optimal value of the corresponding continuous relaxed problem. Many classical algorithms solve the relaxed problem (non-linear or linear programming) as the initial step for solving the given IP problem. It is known that the continuous problem can be solved in polynomial time. The problems considered in this work are benchmarked using the branch \& bound algorithm. The problem with a larger gap $B_1$ (same number of variables) requires more nodes for convergence, highlighting the relevance even for problems with a small number of variables.

A problem is termed tractable if it is solvable. Non-linearity in the problem can make it intractable, for example, a quadratic IP problem is \textit{undecidable} and cannot have any algorithm to solve it \cite{jeroslow1973there}. While solving the non-linear problems, the classical algorithms do not always converge even if the problem is \textit{decidable} due to the presence of many local minima. The common IP solvers struggle to solve NLIP problems. They suffer from inaccurate solutions and large computing times as compared to the ILP counterparts \cite{kronqvist2019review}. In the case of branch \& bound, there is an explosion in the number of nodes for a modest size problem, limiting the size of the NLIP that can be solved classically. The degree of non-linearity is defined as \cite{kronqvist2019review},
\begin{equation}
B_2 = \frac{n}{n_{tot}} \cross 100\%,
\label{B2}
\end{equation}
where $n$ is the number of variables involved in a non-linear term and $n_{tot}$ is the total number of variables. In general, other metrics such as the degree of non-convexity of the optimization landscape can be defined for measuring the complexity. Non-convexity can be captured indirectly by combining $B_1$ and $B_2$, as non-linearity directly contributes to a complex landscape and a large relative relaxation gap can be a consequence of a non-convex hull.

Lastly, discrete density provides a measure of the number of discrete variables present in the problem. When integer values are introduced in a linear programming problem, the complexity class changes from P to NP-hard. Hence for any two integer programming problems, the one with more discrete variables will be harder to solve and the metric discrete density becomes relevant. It is defined as \cite{kronqvist2019review},
\begin{equation}
B_3 = \frac{n_{int} + n_{bin}}{n_{tot}} \cross 100\%,
\label{B3}
\end{equation}
where $n_{int}$ and $n_{bin}$ are the number of integer and binary discrete variables respectively while $n_{tot}$ is the total number of variables. This parameter is useful for categorizing the complexity of different MIP problems.
The prototypical problems shown in Figure~\ref{Tab}(a) are categorized based on the complexity via the metrics discussed in this Section, also represented by the black dots in Figure~\ref{Tab}(b),(c), and (d). The chosen problems ($P_1-P_4$ in Figure~\ref{Tab}(a)) are then solved exactly, and the results are discussed next.

\section{Results}
\label{results}
\begin{figure*}[t]
\centering
\includegraphics[width = 1\textwidth,trim={10cm 0.6cm 10cm 0.7cm},clip]{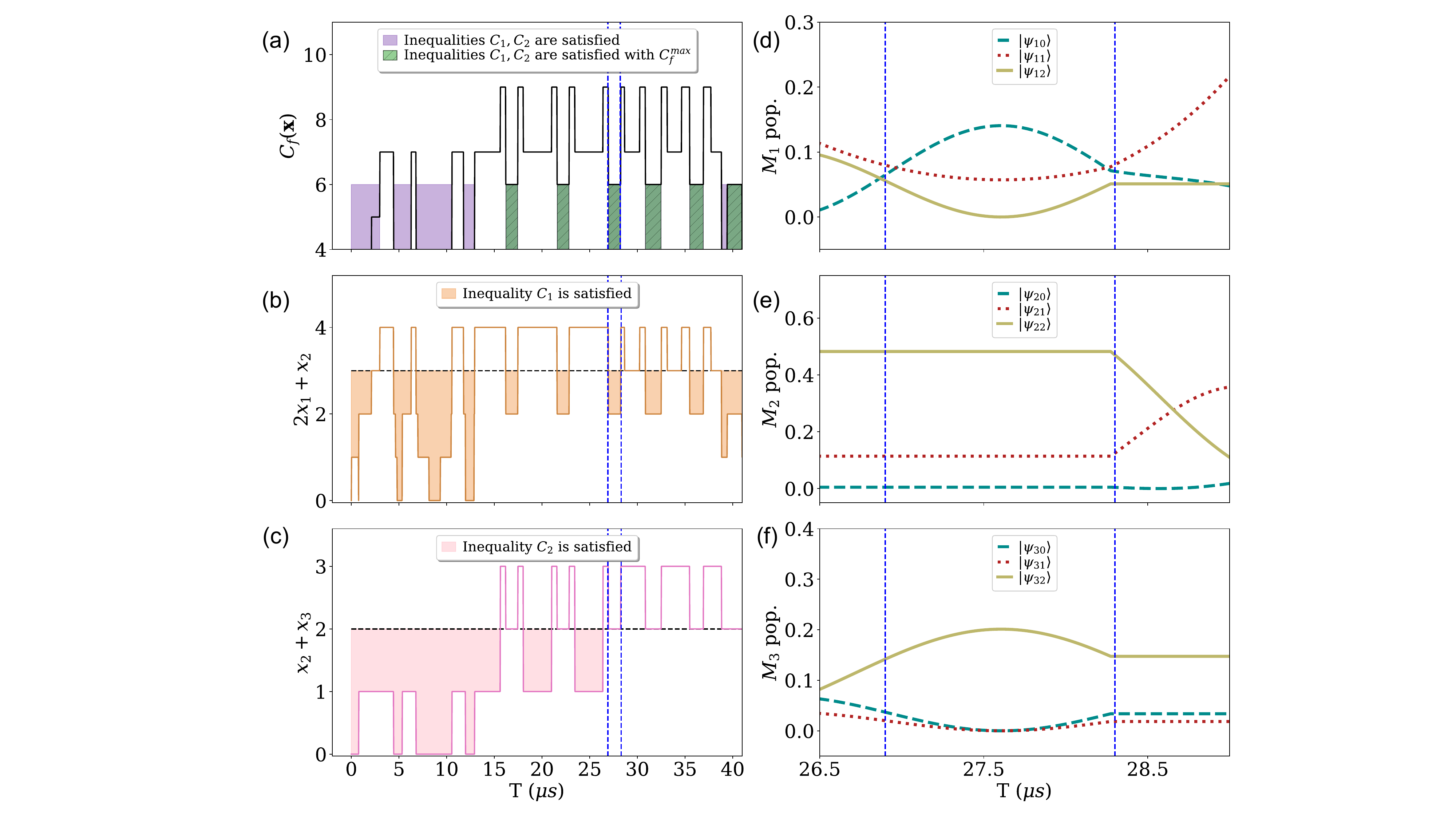}
\caption{\textbf{Integer linear programming problem}: maximize $C_f(\mathbf{x}) = 3x_1 + 2x_2 + x_3$ where constraints are $x_i \in \{0,1,2\}$, $C_1:2x_1+x_2 \leq 3$ and $C_2: x_2 + x_3 \leq 2$. $C_f=6$ is the true solution (global maxima in this case) to the problem calculated by considering all the possible values of the variables. (a) Depicts the value of the cost function of the problem with the shaded region (both in purple and green with slanted lines) corresponding to the time intervals when all the inequalities are satisfied simultaneously. Specifically, the shaded green-slanted-line region represents the time intervals for which the optimal value (maxima) of the cost function is reached while satisfying the constraints. Panels (b-c) show the implementation of the two constraints of the problem, both constraints are implemented using coupling Hamiltonians given by Equations~(\ref{H1eq}-\ref{H2eq}). The shaded regions in (b) and (c) correspond to the time intervals when the constraint inequalities $C_1,C_2$ are individually satisfied respectively. The two vertical blue-dashed lines mark the time interval during which the population of the levels in (d-f) for different manifolds are shown. The state with the highest population assigns the value to the variables which are then used to calculate $C_f, C_1$, and $C_2$. }
\label{QILP1}
\end{figure*}

\begin{figure*}[t]
\centering
\includegraphics[width = 1\textwidth,trim={10cm 0.6cm 10cm 0.5cm},clip]{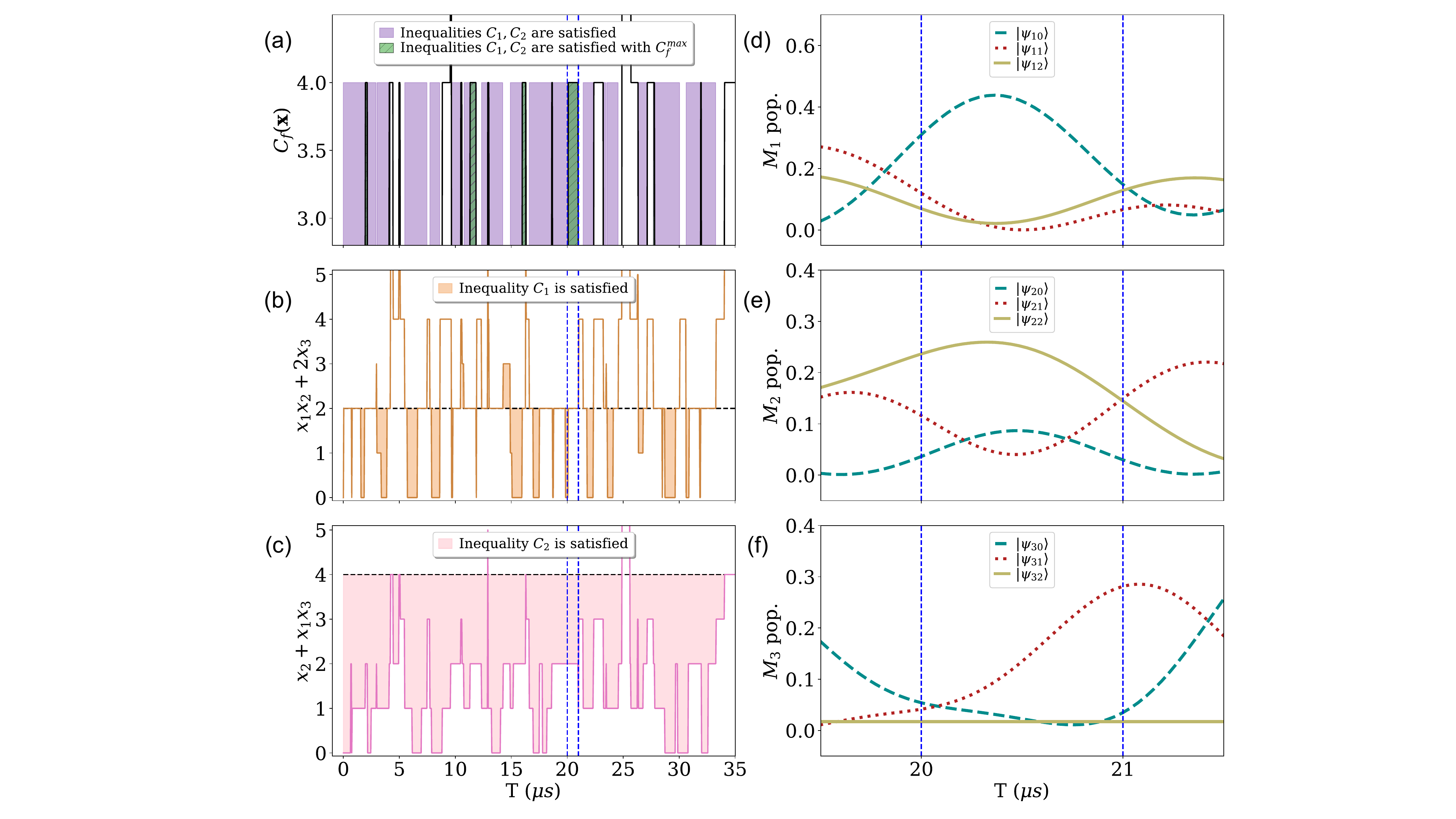}
\caption{\textbf{Non-linear integer programming problem}: maximize $C_f(\mathbf{x}) = 2x_2x_3$ where constraints are $x_i \in \{0,1,2\}$, $C_1:x_1x_2 + 2x_3 \leq 2$ and $C_2: x_2 + x_1x_3 \leq 4$. $C_f=4$ is the true solution (global maxima) to the problem found by considering all the possible values of the variables. In (a) the value of the non-linear cost function of the problem is shown with the shaded region (including the ones with slanted lines) corresponding to the time interval when both inequalities are satisfied simultaneously. Panels (b-c) show the implementation of the constraints of the problem. The shaded regions in (b-c) correspond to the time intervals when the constraint inequalities are satisfied. The population of the levels for different manifolds as shown in (d-f) is during the time interval marked by the two vertical dashed-blue lines.}
\label{QNILP}
\end{figure*}

Figure~\ref{QILP1} demonstrates the working principle of our algorithm where we first consider a simple ILP problem whose optimal solution and the corresponding populations of the states are shown. The chosen problem corresponds to the example IP given by Equation~(\ref{ExP}) containing three integer variables and two linear constraints ($P_1$ from Figure~\ref{Tab}(a)).
The coupling Hamiltonians (Equations~(\ref{H1eq}-\ref{H2eq})) corresponding to the two constraints $C_1, C_2$ are applied consecutively to evolve an initial state in time.
Panel (a) shows the cost function $C_f$ changing in time (solid black line) where the true solution to the problem is given by $C_f=6$. 
The multi-objective function (Equation~\ref{OF}) is minimized to find the maximum of $C_f$ while satisfying both the constraints $C_1, C_2$ shown as the shaded regions in panel (a). 
Time intervals where both constraints are satisfied correspond to the \textit{feasible} regime of the original problem, depicted by the filled regions in panel (a). The maximum value of the cost function in this \textit{feasible} regime is the solution to the problem which is shown as green-shaded regions with slanted lines in panel (a).
Similarly, the individual constraints (solid lines) and their corresponding inequality fulfilled time intervals (shaded) are depicted, $C_1 \leq 3$ in panel (b) and $C_2 \leq 2$ in panel (c).
Panels (d-f) show the populations of the states in manifolds $M={1,2,3}$ for the region between the two vertical dashed (panels a-c) blue lines. The population of the states $\ket{\psi_{10}}$, $\ket{\psi_{22}}$, and $\ket{\psi_{32}}$ in panels (d),(e) and (f) respectively are the highest during the time interval enclosed between the vertical blue lines, assigning the values $x_1 = 0$, $x_2 = 2$, and $x_3 = 2$ respectively. In this way the cost function $C_f$ and the constraints $C_1, C_2$ are then calculated by decoding the variables by the measurement of the populations. Next, we consider a more complex case as an example problem.

Figure~\ref{QNILP} shows the flexibility of our algorithm by solving a problem with a non-linear cost function and constraints. 
The sample integer programming problem is intentionally chosen to test our scheme on a \textit{harder} problem, refer to Equation~(\ref{B2}) ($P_2$ in Figure \ref{Tab}(a)).
In Figure \ref{QNILP}, panel (a) shows the exact optimal solution of the problem given at times marked by the shaded green region (with slanted lines) while panels (b) and (c) correspond to the individual constraints. 
The protocol scheme follows the same steps as described for the previous problem. 
Although the optimal solution is reached at $T \sim 2 \mu s$ as shown by the earliest green (slanted lines) region in panel (a), however, we chose to show the longest green (with lines) interval for better understanding. The population of the states during the interval bounded by the two dashed-blue lines are shown in panels (d-f). The population of the states gives the value of the variables $x_1 = 0$ ($\ket{\psi_{10}}$ in panel (d)), $x_2 = 2$ ($\ket{\psi_{22}}$ in panel (e)) and $x_3 = 2$ ($\ket{\psi_{32}}$ in panel (f)) that are used to calculate the cost function that is $C_f = 4$. For constraint $C_2$, any state of any manifold can be populated, while for $C_1$, $x_3=2$ is forbidden, translating to state $\ket{\psi_{32}}$ not being coupled to any other state.   

\begin{figure*}[t]
\centering
\includegraphics[width = 0.6\textwidth,trim={2cm 2cm 0cm 2cm},clip]{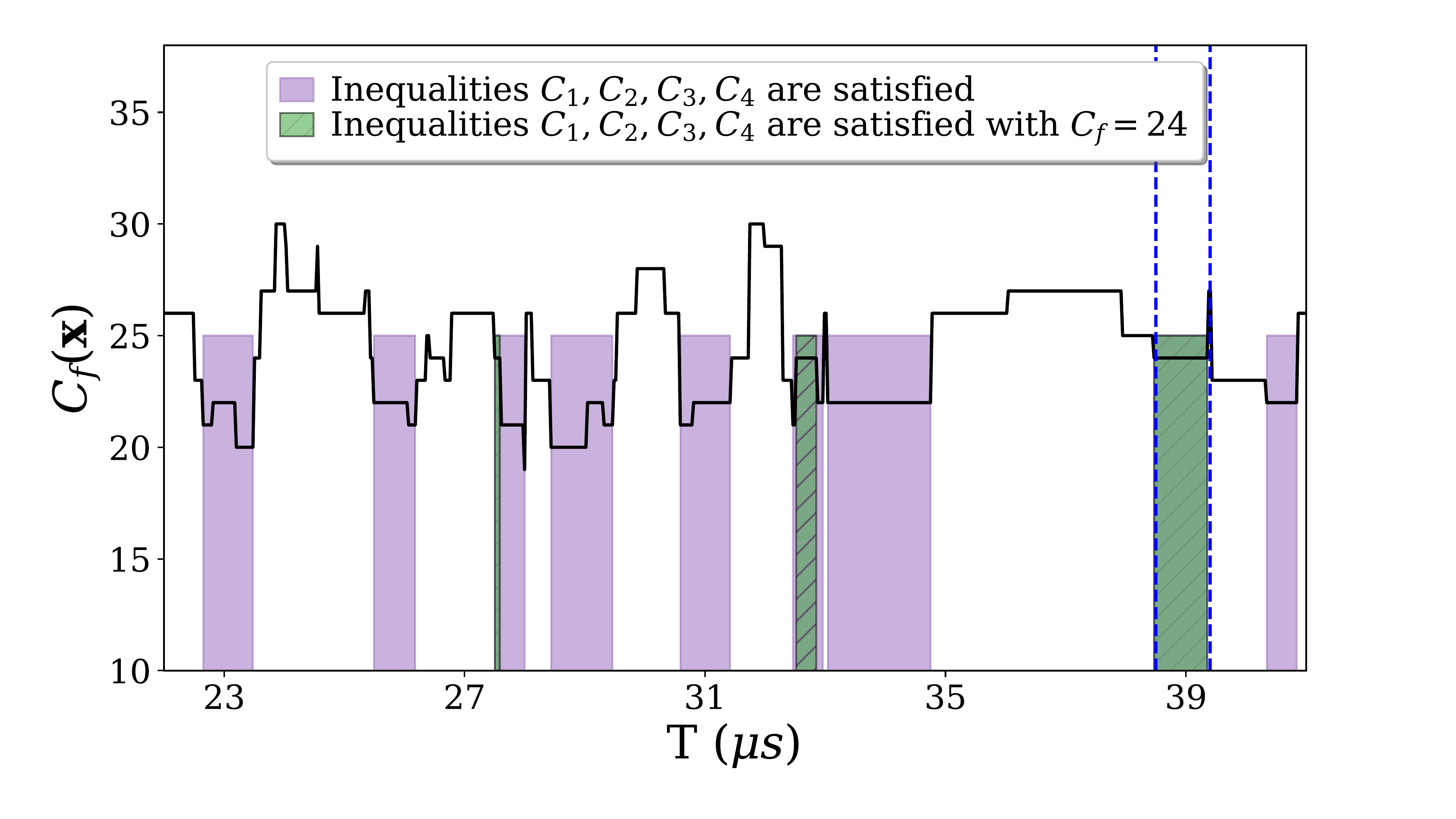}
\caption{\textbf{Integer linear programming problem}: maximize $C_f(\mathbf{x}) = x_1+2x_2+x_3+4x_4+3x_5+x_6+2x_7+5x_8$ where constraints are $x_i \in \{0,1,2\}$, $C_1:x_1+x_2+x_3 \leq 1$, $C_2: 2x_4 + 3x_8 \geq 5$, $C_3:2x_5+2x_6 + x_7 \leq 3$, and $C_4: x_3 + x_7 \leq 1$. $C_f=25$ is the solution to the problem calculated using a brute force approach. (a) shows the cost function for the problem $P_3$ (in Figure \ref{Tab}(a)). The shaded region corresponds to satisfying all the inequalities, and the green region (with slanted lines) shows a near-optimal solution.}
\label{QILP2}
\end{figure*}

The next IP problem considered in Figure~\ref{QILP2} consists of eight variables and four constraints. The cost function is shown in Figure~\ref{QILP2}, where the green regions (with slanted lines) correspond to the near-optimal solution $C_f=24$, and the true solution is $C_f=25$. 
The chosen problem, $P_3$ in Figure \ref{Tab}(a), is more complicated as compared to the previous examples since the complexity of the IP problem increases exponentially with the number of discrete variables $B_3$ (depicted in Figure \ref{Tab}(d)). This complexity results in the frequency and the width of the green-slanted-line regions for solving $P_3$ (Figure~\ref{QILP2}) being less as compared to a simpler problem of three variables $P_1$ (Figure~\ref{QILP1}(a)). However, the near-optimal solution is reached in $40 \mu s$, and the system remains in this configuration for more than $1 \mu s$. We also reach the true solution $C_f=25$ with a different set of parameters that lasts for $0.1 \mu s$ (not shown here) for which the states stay in the optimal configuration. The shaded regions correspond to the times when all the inequalities are satisfied and the cost function during those time intervals has many sub-optimal solutions. For instance, during $T \sim 32-35 \mu s$, the cost function value $C_f=22$ is close to the optimal solution $C_f=25$. 

\begin{figure*}[t]
\centering
\includegraphics[width = 1.0\textwidth,trim={0cm 5cm 2cm 5.5cm},clip]{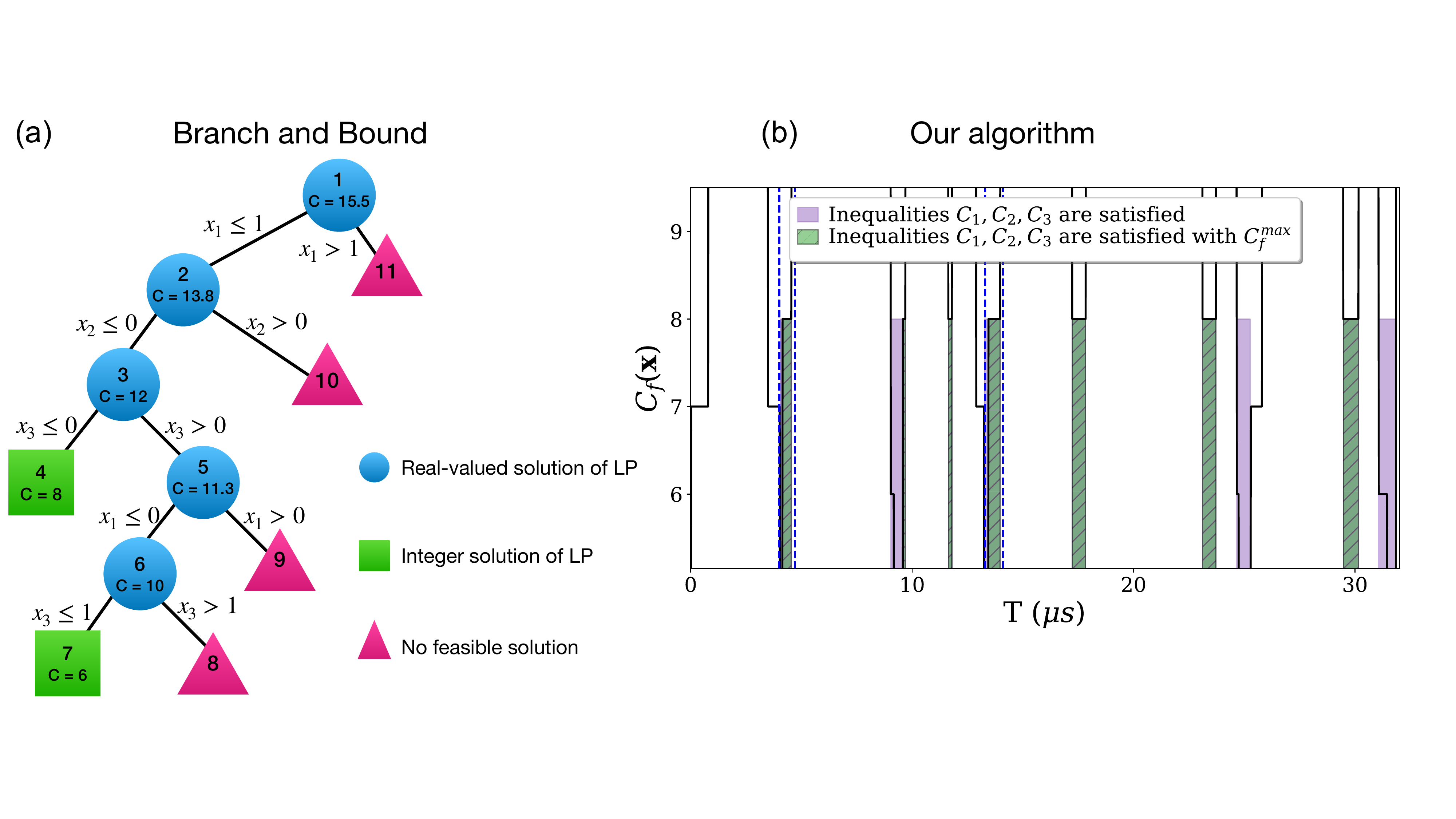}
\caption{\textbf{Integer linear programming problem}: max $C_f(\mathbf{x}) = 8x_1 + 7x_2 + 6x_3$ where constraints are $x_i \in \{0,1,2\}$, $C_1:3x_1+x_2 + 3x_3 \leq 5$, $C_2: x_1+ 2x_2 + x_3 \leq 3$ and $C_3: 4x_2 + x_3 \leq 2$ with $C_f=8$ being the solution (global maxima). (a) shows the \textbf{classical branch and bound} method for solving the problem and (b) shows the cost function value and the optimal solution using our algorithm. The regions between the dotted (vertical) lines correspond to the solution to the problem for extended time intervals ($\sim 1\mu s$).}
\label{QILPB}
\end{figure*}

To benchmark the performance of our algorithm, an implementation of branch \& bound (B\&B) is considered which is shown in Figure~\ref{QILPB}. 
The classical B\&B is used to solve the problem $P_4$ from Figure~\ref{Tab}(a), which required $11$ nodes (iterations) to converge to the optimal solution. Panel (a) in Figure~\ref{QILPB} shows the explicit nodes involved in the B\&B algorithm. Circular blue nodes are intermediate steps, where the problem branches into sub-problems with additional constraints on the variables. B\&B relies upon solving the relaxed problem and bounding the solution space by constraining the problem. 
The IP problem is relaxed to have continuous variables, and the solution of the continuous problem provides an upper bound for the solution of IP. A lower bound is found by assigning integer values to the optimal continuous variables ($v_1,...,v_n$). The continuous variable (e.g., $x_i$) with the largest fractional part is chosen to branch and solve the problem at two nodes, such that Node 1: $x_i \leq \floor{v_i}$ and Node 2: $x_i > \floor{v_i}$. In panel (a), the upper bound of the cost function at each node is given by the value $C$ which kept decreasing after the subsequent branching, and the lower bound did not change for the problem $P_4$. The complexity of $P_4$ is captured by the high $B_1 = 93\% > 50\%$ value \cite{kronqvist2019review}, corresponding to the large difference in the solution of the relaxed continuous problem and the discrete IP. For comparison, problem $P_1$ has $B_1 =8.3\%$ and it converges after $3$ nodes in B\&B to the optimal solution. In general, if the value of $B_1$ is higher for a problem, it takes more branches for the upper bound to converge to the integer solution.
The branching can be terminated in three ways, (1) the optimal solution contains all the integer variables as shown by green squares in panel (a), (2) the relaxed problem becomes infeasible under the constraints corresponding to the pink triangles, and (3) the stopping criteria of either the maximum number of nodes or the maximum time are reached, which is required when the problem has a high relative continuous relaxation gap. Many state-of-the-art solvers use one of the implementations of the B\&B algorithm, hence they also suffer when one of the parameters given in Equations~(\ref{B1}),~(\ref{B2}),~(\ref{B3}) becomes larger for different instances of the problem.
The algorithm for the given problem $P_4$ terminates after $11$ nodes. The cost function for the nodes containing integer solution of LP (values in green squares) is compared and the optimal solution corresponds to the maximum of them. For $P_4$, the optimal solution is $(x_1,x_2,x_3)=(1,0,0)$ with cost function value $C=8$. Panel (b) shows the result of our algorithm, in which the solution is reached as early as $5\mu s$ during the first layer of the Hamiltonians.
For problems with a larger number of variables, alternate hybrid approaches can be used to potentially provide more accurate results compared to currently available classical algorithms. In such a hybrid quantum-classical scheme, B\&B can be used to construct the tree and our quantum algorithm can solve the constricted problem at each node of the tree.        

\section{Conclusions and Outlook}
\label{Conclusion}
Implementation of most of the algorithms on a quantum system requires a large number of qubits and they rely upon the quantum entanglement of a many-body system to gain an advantage over classical methods. Alternatively, an advantage can potentially be achieved by exploiting the superposition principle. This work provides another approach for obtaining a quantum advantage over classical algorithms without explicitly using entanglement.
Generally, combinatorial optimization problems can be formulated mathematically either as a QUBO or an IP problem. 
The architecture of the current quantum devices makes QUBO formulation a more popular choice for executing algorithms, however, IP is widely used to formulate industrial optimization problems due to it being more efficient in terms of the number of variables. 
To solve an IP problem on a quantum device, QUBO is used as a stepping stone, which is an indirect approach, thus increasing the number of qubits required. 
In the current literature for solving IP on a quantum system, \cite{khosravi2023mixed} considers a problem with three binary variables with a probability $85-90\%$ for finding the solution. While in \cite{zhao2022hybrid}, a problem with two discrete variables is encoded with a solution not better than the classical counterpart of the algorithm. In \cite{bernal2020integer}, a linear graph of two nodes for solving a minimum dominating set problem required 5 qubits to get a solution with probability $90-95\%$.

Our approach for the first time, introduces a non-QUBO-based algorithm by directly mapping the IP problem to a quantum system and solving it by exploiting the superposition principle. It opens up a new pathway of encoding and algorithmic processing within a single atom. The algorithm we provide can handle both linear and non-linear IP problems while it only requires a single atom to implement it.  
We chose a few prototype problems with varying complexities characterized by the benchmarking metrics namely, relative continuous relaxation gap, degree of non-linearity, discrete density, and size of the problem. The problems are then encoded to a multi-level Rydberg atom with each manifold of states representing integer variables. The manifolds are probed and selectively populated by optimizing the coupling strength between the different levels for implementing the problem's constraints. The population of the states is then measured to decode the optimal values of the integer variables to find the solution of the problem which is reached in microseconds $\mu s$ with high accuracy. A benchmark is performed with the classical B\&B and our algorithm for a problem with a large relative continuous relaxation gap. The direct algorithm solved the problem in a one-iteration while the B\&B algorithm needed $11$ iterations to converge, showcasing the potential of our algorithm for a hybrid approach. The classical algorithms suffer more in terms of resources and accuracy for solving non-linear problems as compared to the linear problems \cite{kronqvist2019review}.
    
There are a few ways in which a larger problem can be solved under our framework, which is the outlook for this work. One atom is used in our algorithm to solve a problem with $8$ variables, if we use hundreds of atoms, in principle that can encode a problem with thousands of variables. The constraints can be implemented by mediating the interaction between different atoms. All the computations can then be performed in parallel thereby decreasing the time for solving the large problem. Another possible approach would be to divide the larger IP problem into sub-problems using the Benders decomposition \cite{rahmaniani2017benders}. But instead of solving the relaxed continuous sub-problems using classical B\&B to bound the solution space, we can provide exact solutions by solving the integer (without relaxation) sub-problems on the individual atoms in parallel. This can provide better approximate solutions and enhance the bounds provided by the branch and bound method. For comparison, the state-of-the-art classical solver (BARON), solves a non-convex mixed non-linear IP problem with $2500$ variable in $300 s$ \cite{kronqvist2019review}. In this way, we have the perspective of solving problems where the current classical algorithms struggle. 

\ack
This work is funded by the German Federal Ministry of Education and Research within the funding program “Quantum Technologies - from basic research to market” under Contract No. 13N16138.

\bibliographystyle{iopart-num.bst}
\bibliography{IP}

\end{document}